\documentclass[aps,prc,twocolumn,showpacs,groupedaddress,preprintnumbers,amsmath,amssymb,floatfix]{revtex4}
\usepackage[dvipdfm]{graphicx}
\begin{document}
%===============================%
%===%        AUTHORS        %===%
%===============================%
\author{M. Krmar and N. Jovan\v{c}evi\'c}
\affiliation{
University of Novi Sad, Faculty of Science, Department of Physics, Trg Dositeja Obradovi\'ca 4, 21000 Novi Sad, Serbia
}
\author{D.~Nikoli\'c}
\affiliation{
National Institute for Nanotechnology, Edmonton AB, T6G 2M9, Canada
}
%===============================%
%===%         TITLE         %===%
%===============================%
\title{Gamma ray production in inelastic scattering of neutrons\\
produced by cosmic muons in $^{56}$Fe}
%===============================%
%===%         DATUM         %===%
%===============================%
\date{\today}
%===============================%
%===%        ABSTRACT       %===%
%===============================%
\begin{abstract}
We report on the study of the intensities of several gamma lines emitted after the inelastic scattering of neutrons in
$^{56}$Fe. Neutrons were produced by cosmic muons passing the 20t massive iron cube placed at the Earth's surface
and used as a passive shield for the HPGe detector. Relative intensities of detected gamma lines are compared with the results
collected in the same iron shield by the use of $^{252}$Cf neutrons. Assessment against the published data from
neutron scattering experiments at energies up to 14~MeV is also provided.
\end{abstract}

%===============================%
%===%     MAKE PACS         %===%
%===============================%
\pacs{23.20.Lv,98.70.Vc,25.85.Ec,25.30.Mr}
%===============================%
%===%     MAKE TITLE        %===%
%===============================%
\maketitle
%===============================%
%===%   SEC: INTRODUCTION   %===%
%===============================%
\section{Introduction}
\label{Sec:Introduction}

The possible exploration of chemical composition of planets and extraterrestrial materials
\cite{Reedy:1973} by analysis of gamma spectra emitted from the surface, renewed the interest
for gamma ray production cross sections in stable nuclei \cite{Boynton:2007,Skidmore:2009}.
The energy level structure of the stable $^{56}\rm{Fe}$ has been studied extensively in the past
via {\em inelastic neutron scattering} (INS) \cite{Kiehn:1954,Day:1956,Cranberg:1956,Bredin:1964,
Rogers:1972,Mellema:1986,Pedroni:1988}, and pertinent cross sections for the gamma ray production exist
\cite{Rogers:1972,Savin:1976}. However, secondary neutrons produced by primary cosmic rays have much
higher energies compared to reactor ones and cross sections for gamma ray production by the use of
neutrons with energies up to 150~MeV were studied recently \cite{Sisterson:2006,Castaneda:2007}.

Despite $^{56}\rm{Fe}(n,n'\gamma)$ nuclear reaction being well-explored, few publications
\cite{Lachkar:1974,Xiamin:1982} report on relative intensities of gamma lines induced by
INS on $^{56}\rm{Fe}$ nuclei. Pre-World~War~II iron is widely used as a passive shield and
846.8~keV gamma line, induced by inelastic scattering of neutrons produced by cosmic muons,
is a well-known feature in background spectra \cite{Heusser:1996}. Even though detailed knowledge
of all possible sources of radiation could be of crucial importance in numerous low-background
experiments, the gamma radiation due to excitation of $^{56}\rm{Fe}$ nuclei by inelastic scattering
of muon-produced neutrons received little attention.

Reliable experimental studies \cite{Rogers:1972,Savin:1976} of cross sections for $^{56}\rm{Fe}(n,n'\gamma)$
reaction have indicated that detected intensities of gamma lines depend on incident neutron energy, and it is
to be expected that the same holds for {\em relative} intensities. Indeed, for incident neutron energies between
6.3~MeV and 14.2~MeV, relative intensity of 1238.3~keV gamma line changes by almost 80\% \cite{Lachkar:1974,Xiamin:1982}.
There are several muon-stimulated mechanisms of neutron production with significant contributions to the
background \cite{Kudryavtsev:2003}: muon capture, muon-induced spallation reactions, muon-induced hadrons cascades,
and photonuclear reactions (high energy photons from a muon-induced electromagnetic cascade).
At the Earth's surface and shallow depths, the dominant neutron production mode is negative muon capture.
This process occurs by stopping muons and plays a significant role \cite{Galbiati:2005} because the mean
energy of surface muons is about 4~GeV. On the other hand, the energies of neutrons emitted after muon capture do not exceed several MeV
\cite{Schroeder:1974}. High energy muons, also present at Earth's surface in much lower arrival numbers, penetrate
to the larger depths. It has been demonstrated through several background measurements in underground laboratories
\cite{Kudryavtsev:2003,Galbiati:2005,Agafonova:2009} that the hard muon component may create neutrons with energies
up to several hundred MeV.

The goal of this paper is to identify characteristic $^{56}\rm{Fe}$ gamma lines in background spectra registered by a
iron-shielded HPGe detector. The relative intensities of measured gamma transitions are assessed against published results
on the same gamma transitions detected in neutron beam scattering experiments. We use $^{252}\rm{Cf}$ source to get
the qualitative information about the average energy of muon-created neutrons in the iron shield and to infer relative
intensities of gamma lines associated with scattering of muon-induced neutrons and fission neutrons.
%===============================%
%===%      SEC: SETUP       %===%
%===============================%
\section{Experimental Setup}
\label{Sec:ExperimentalSetup}

The HPGe detector used in experiments is located in the low-background laboratory of the Department
of Physics in Novi Sad (80~m~amsl). The detector is placed inside a cube chamber made of pre-World~War~II iron with a useful
cube-shaped inner volume of $1 \rm{m}^3$. Iron walls are 25~cm thick and the total mass of the shield is about 20~tons.
No other passive shields were used throughout the measurements. The distance between the top of the iron shield and
30~cm thick concrete roof construction is about 3~m. Iron chamber is placed on concrete base over 2~m away from
walls made of bricks and wood. Relative efficiency of HPGe detector is 25\% with the active detection volume
of Ge crystal being located in the center of the inner volume of the iron cube chamber.
The integral background count rate for energies ranging from 20~keV to 2~MeV was about 1 count per second.

To get experimental evidence about relative intensities of gamma lines following de-excitation of $^{56}\rm{Fe}$ nuclei
after inelastic scattering of fission neutrons, $^{252}\rm{Cf}$ source (having stable neutron emission rate of
$4.5\cdot10^{3} \rm{s}^{-1}$ into $4\pi$~sr) was placed at several different positions inside the iron chamber.
To attenuate gamma radiation originating from $^{252}\rm{Cf}$ fission products accumulated in source, solid angle of
detector as seen from the californium source was screened by a layer of iron. Measurements were done with uncovered
$^{252}\rm{Cf}$ source, as well as with the source packed in 2~cm of paraffin.

%===============================%
%===%   SEC: RESULTS        %===%
%===============================%
\section{Measurements and Results}
\label{Sec:MeasurementsAndResults}

\begin{figure}[bhtp]
\includegraphics[scale=1.0]{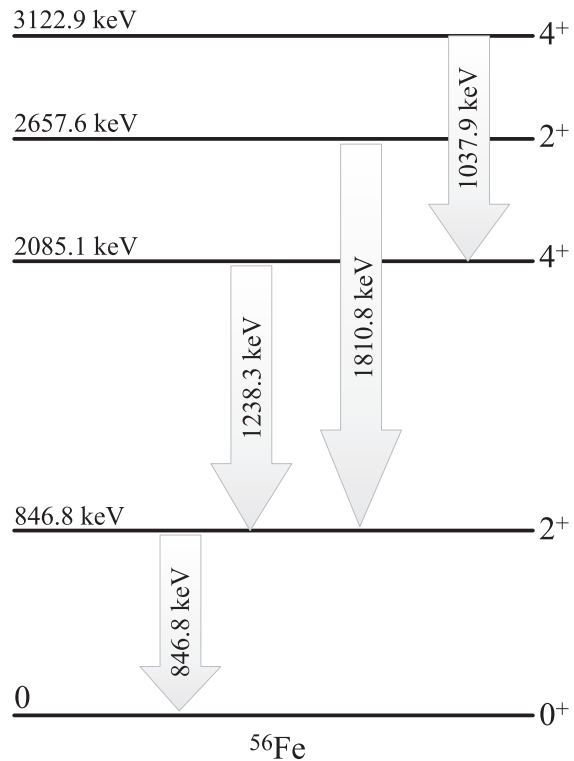}
\caption{\label{Fig:1}
Part of the energy level scheme of $^{56}\rm{Fe}$ with gamma-ray lines investigated in this work.}
\end{figure}

In order to explore the gamma radiation produced by the interaction of neutrons with surrounding iron, long time spectra
was obtained by the summing a number of background spectra, thus amounting to a total observation time of $5.1\cdot10^{6} \rm{s}$.
The strongest gamma transition in Fig.~\ref{Fig:1} is due to 846.8~keV background line with the measured intensity of
$1.16(8)\cdot10^{-3} \rm{s}^{-1}$. Intensities of the remaining three gamma lines in Fig.~\ref{Fig:1} have been normalized
to that of 846.8~keV line and are summarized in Table~\ref{Tab:1}.
Variations in detection efficiency, due to different penetration abilities of gamma photons in iron and detector itself,
were accounted for at each observed energy with the help of GEANT4 simulations \cite{GEANT4}.
In~the first approximation, it was considered that gamma photons are emitted uniformly from a complete volume of the
iron shield. However, simulations also take into account attenuation of the emitted radiation originating from the
part of the iron shield that is shaded by the Dewar vessel.

It is interesting to note that $^{56}\rm{Fe}(4^{+},2^{+}\gamma)$ gamma transition in Fig.~\ref{Fig:1} occurs at energy
that coincides with the background line of $^{214}\rm{Bi}$ ($^{238}\rm{U}$ series member) at 1238.1~keV. Hence, intensity
of $^{56}\rm{Fe}$ 1238.3~keV gamma line was inferred by proper decoupling and subtraction of $^{214}\rm{Bi}$ contribution.
Namely, nine $^{214}\rm{Bi}$ gamma lines in the energy interval between 609.3~keV and 1764.6~keV were used for proper
efficiency calibration of our detector and for deconvolution of the 1238.1~keV $^{214}\rm{Bi}$ portion from the
nearby $^{56}\rm{Fe}$ 1238.3~keV line profile.
We further note that isolated $^{56}\rm{Fe}$~1810.8~keV gamma line appeared
relatively weak in background spectra, whereas another line of interest at 1037.9~keV remained absent. It is the flat
part of the background spectra at the place of the missing 1037.9~keV gamma line that was used in resolving the upper limit
for its intensity reported in Table~\ref{Tab:1}.
%===============================%
%===%       TABLE 1:        %===%
%===============================%
\begingroup
\squeezetable
\begin{table}[thbp]
\caption{\label{Tab:1}
Relative intensities (in \%) of gamma transitions depopulating excited states in $^{56}\rm{Fe}$. Gamma lines
were detected both in the background and by the use of californium source (see text for details).}
    \begin{ruledtabular}
        \begin{tabular}{ccccc}
               & \multicolumn{3}{c}{iron chamber}                                & $^{56}\rm{Mn}$ decay \\
\cline{2-4}
$E_{\gamma}$ (keV)  & background & $^{252}\rm{Cf}$ in air & $^{252}\rm{Cf}$ in paraffin &               \\
\colrule
$1037.9$            & $< 1.4$    & $< 0.2  $       & $< 0.9$                     &      $0.040$         \\
$1238.3$            & $ 8(6)$    & $9.3(7) $       & $ 7(3)$                     &      $0.10 $         \\
$1810.8$            & $11(7)$    & $6.1(14)$       & $ 5(3)$                     &      $27.7 $         \\
        \end{tabular}
    \end{ruledtabular}
\end{table}
\endgroup

Similar procedure was repeated in the case of spectra recorded with $^{252}\rm{Cf}$ source located inside of the iron chamber.
Namely, two types of additional measurements were performed: one with uncovered californium source and the other with
the source enclosed in 2~cm thick paraffin container. Paraffin moderator was used primarily to reduce the mean energy
of fission neutrons emitted by $^{252}\rm{Cf}$. The outcome of these measurements showed that relative intensities
of $^{56}\rm{Fe}$ gamma lines originating after inelastic neutron scattering do not depend on $^{252}\rm{Cf}$ source
position in the chamber. Characteristic times of both measurements were about $6\cdot10^{4} \rm{s}$. In this relatively
short period of time, intensities of background $^{214}\rm{Bi}$ lines were much lower than the intensity of lines emitted
from $^{56}\rm{Fe}$ excited nuclei and there was no need to subtract a possible contribution due to 1238.1~keV
background line. Relative intensities of all three gamma lines of interest obtained in these measurements are also
presented in Table~\ref{Tab:1}.
%%%===================================%
%%%===         FIG. 1             %===%
%%%===================================%
%%%===================================%
Due to slow-down of neutrons in the paraffin, intensities of the observed gamma lines became
significantly lower, however, relative intensities remained similar to those obtained with uncovered $^{252}\rm{Cf}$ source.

\begin{figure}[tbhp]
\includegraphics[scale=1.0]{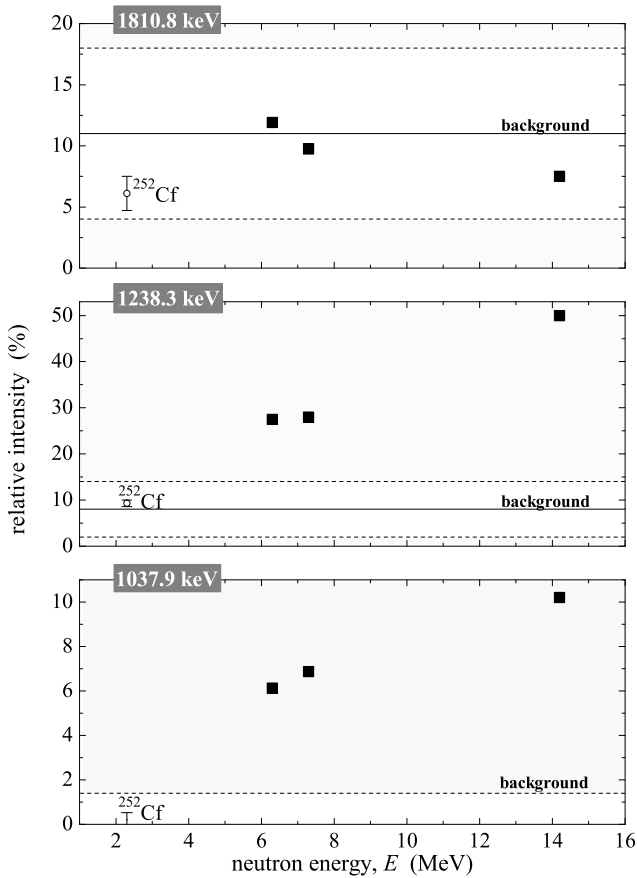}
\caption{\label{Fig:2}
Relative intensities of gamma lines detected both in the background spectra and the spectra measured by the use of
$^{252}\rm{Cf}$ source. Results of pertinent gamma transitions obtained in neutron beam measurements at
6.3~MeV, 7.3~MeV, and 14.2~MeV \cite{Lachkar:1974,Xiamin:1982} are shown in solid squares (see text for details).}
\end{figure}

Figure~\ref{Fig:2} juxtaposes the relative intensities of 1037.9~keV, 1238.3~keV, and 1810.8~keV gamma lines
measured in our iron shield with those from neutron beam experiments. The solid squares
denote results of relative intensities of gamma lines obtained in the standard neutron beam experiments. Data for 6.3~MeV and
7.3~MeV neutrons were published by Lachkar~{\em et al}~\cite{Lachkar:1974} and 14.2~MeV data are due to Xiamin~{\em et al}~\cite{Xiamin:1982}.
Our result acquired by the use of uncovered californium source is also given in Fig.~\ref{Fig:2} at the adopted
mean energy of 2.3~MeV for $^{252}\rm{Cf}$ fission neutrons. The solid horizontal lines in Fig.~\ref{Fig:2} represent the values of
the relative intensities of gamma lines measured in the background spectra whereas the dashed lines denote a 1-$\sigma$
uncertainty corridor.
%%%===================================%
%%%===         FIG. 2             %===%
%%%===================================%
%%%===================================%
Another possible source of gamma radiation emitted from the excited $^{56}\rm{Fe}$ nucleus is the
${}^{56}\rm{Fe}(\mu,\nu_{\mu}){}^{56}\rm{Mn}$ reaction where muon capture occurs without the emission of neutrons.
Heisinger~{\em et al}~\cite{Heisinger:2002} estimated that about 15.7\% of the overall number of muon captures in iron
take place through this channel. The product of the muon capture, ${}^{56}\rm{Mn}$, decays to the $^{56}\rm{Fe}$ and
we report the relative intensities of three gamma lines of interest in the last column of Table~\ref{Tab:1}.
Very weak 1238.3~keV gamma radiation emitted in the decay of ${}^{56}\rm{Mn}$ suggests that almost all 1238.3~keV photons
come from INS at ${}^{56}\rm{Fe}$ nuclei.
Relatively strong gamma line of 1810.8~keV is emitted in decay of ${}^{56}\rm{Mn}$ and it is possible that both processes,
inelastic neutron scattering and decay of ${}^{56}\rm{Mn}$, contribute jointly to the total count rate under 1810.8~keV
gamma peak in the background spectra. The evidence about possible ${}^{56}\rm{Mn}$ involvement in measured gamma lines in
the background spectra can be obtained by analysis of its high energy part. One of the strongest gamma lines in decay
scheme of ${}^{56}\rm{Mn}$ has 14.3\% intensity and comes from the 2113.2~keV transition depopulating the 2959.9~keV level.

The neutrons created by cosmic muons have insufficient energy to excite the 3122.9~keV level in ${}^{56}\rm{Fe}$, and
1037.9~keV gamma line remains unobserved in the background spectra. One can draw similar conclusion for radiative
depopulation of nearby 2959.9~keV level (not shown in Fig.~\ref{Fig:2}). Recent measurements by
Castaneda~{\em et al}~\cite{Castaneda:2007} showed that cross sections for neutron induced 1037.9~keV and 2113.2~keV
gamma ray production from iron have similar values in broad energy range.
The cross sections for both transitions appear much lower compared to the cross section for production of 846.8~keV
gamma transition. For example, the cross section for 846.8~keV gamma transition at the neutron energy of 6.5~MeV
is almost fifty times larger than the cross section for the 1037.3~keV and the 2113.2~keV transitions. In this case, the 2113.2~keV
gamma line in the background spectra should originate from the decay of ${}^{56}\rm{Mn}$ alone.
Unfortunately, the upper limit of the background spectra used in this study was 2~MeV. Thus, we had new background data
accumulated in the extended range of energies with an upper limit slightly over the ${}^{208}\rm{Tl}$ 2614.5~keV gamma line,
used here to calibrate the high-energy part of spectrum. However, this new spectra accrued for $1.04\cdot10^{6} \rm{s}$
also contained no gamma lines at 2113.2~keV. Flat part of continuum, at the place where the 2113.2~keV gamma line should have appeared,
can serve as an estimate of possible influence that ${}^{56}\rm{Mn}$ decay may exert on the overall intensity of
the 846.8~keV ${}^{56}\rm{Fe}$ gamma line.

%===============================%
%===%   SEC: DISCUSSION     %===%
%===============================%
\section{Discussion and conclusions}
\label{Sec:DiscussionAndConclusions}
Our measurements show that inelastic scattering of neutrons produced by cosmic muons in ${}^{56}\rm{Fe}$ can create only
three gamma lines: relatively strong 846.8~keV and pair of weak ones, 1238.3~keV and 1810.8~keV. Muon-induced neutrons,
as well as fission neutrons, do not produce 1037.9~keV gamma line as this transition de-excites poorly populated 3122.9~keV
level. Both types of neutrons, fission as well as cosmic, possess continuous energy distribution. Absence of 1037.9~keV
gamma line, both in the background spectra and the spectra collected by the use of californium source, clearly suggest that number
of neutrons in the high-energy tail above the 3122.9~keV of both energy distributions is insufficient to produce a measurable
number of 1037.9~keV photons. Contrary to our findings, the previously published results shown in Fig.~\ref{Fig:2} and
obtained by the use of beams of neutrons at 6.3~MeV, 7.3~MeV, and 14.2~MeV detected up to 10\% relative intensity for 1037.9~keV
gamma line.

Data presented in Fig.~\ref{Fig:2} for 1238.3~keV gamma line show an apparent increase of relative intensities
with different neutron energies between 6.3~MeV and 14.2~MeV, a trend that can be explained in terms of results
reported by Savin~{\em et al}~\cite{Savin:1976}. Namely, the cross section for emission of 846.8~keV normalization line
increases at energies below 3.5~MeV and remains independent of neutron energy at the higher incident
energies. On the other hand, cross section for emission of 1238.3~keV photons uniformly increases with the energy
of incident neutrons. The energy threshold for emission of 846.8~keV gamma photons is significantly lower than energy
threshold for 1238.3~keV line. Hence, relative intensity of 1238.3~keV gamma line measured in the background spectra
(incapable of exciting 3122.9~keV level) should be lower than relative intensity measured with neutrons having
energies of 6.3~MeV and higher. Results presented in Fig.~\ref{Fig:2} support this argument.

Relative intensities of 1810.8~keV gamma line show just opposite tendency from 1238.3~keV case, as can be seen
in Fig.~\ref{Fig:2}. It can be explained once again in terms of experimental results of Savin~{\em et al}~\cite{Savin:1976}:
the cross section for production of 1810.8~keV photons has maximum at about 4~MeV and at higher neutron energies
uniformly decreases. Unfortunately, in the background spectra of the present study 1810.8~keV gamma line appears very weak
and relative intensities were determined with large statistical uncertainties.
Obtained values for relative intensity of 1238.3~keV gamma line, measured both in the background spectra and the spectra
induced by a californium source, are mutually compatible; in both spectra the 1037.9~keV gamma line is missing
and relative intensities of 1810.8~keV gamma lines appear consistent within error margins.
Comparing the results of Mannhart~\cite{Mannhart:1987} and Schr\"oder~{\em et al}~\cite{Schroeder:1974}, one finds
that both fission neutrons and muon-induced neutrons obey similar energy distributions. Further on, compatibility
of our measured neutron energy spectra obtained with and without $^{252}\rm{Cf}$ source suggest that the dominant
mechanism for neutron production at the Earth's surface is due to muon-capture.
To get better experimental evidence about possible influence of ${}^{56}\rm{Mn}$ decay on intensities of gamma
lines emitted after neutron scattering in natural iron, intensity of 2113.2~keV gamma line should be acquired in
long-time background spectra.
%===============================%
%===%   ACKNOWLEDGMENTS     %===%
%===============================%
\vspace*{5mm}
\section*{Acknowledgments}
\vspace*{-5mm}
The authors acknowledge financial support from the Ministry of Science and Technological Development
of the Republic of Serbia, under the Project No.~141002 (Nuclear Spectroscopy and Rare Processes).
%===============================%
%===%      BIBLIOGRAPHY     %===%
%===============================%
%%\bibliography{Nikolic_Biblography}

\begin{thebibliography}{23}
\expandafter\ifx\csname natexlab\endcsname\relax\def\natexlab#1{#1}\fi
\expandafter\ifx\csname bibnamefont\endcsname\relax
  \def\bibnamefont#1{#1}\fi
\expandafter\ifx\csname bibfnamefont\endcsname\relax
  \def\bibfnamefont#1{#1}\fi
\expandafter\ifx\csname citenamefont\endcsname\relax
  \def\citenamefont#1{#1}\fi
\expandafter\ifx\csname url\endcsname\relax
  \def\url#1{\texttt{#1}}\fi
\expandafter\ifx\csname urlprefix\endcsname\relax\def\urlprefix{URL }\fi
\providecommand{\bibinfo}[2]{#2}
\providecommand{\eprint}[2][]{\url{#2}}

\bibitem[{\citenamefont{Reedy et~al.}(1973)\citenamefont{Reedy, Arnold, and
  Trombka}}]{Reedy:1973}
\bibinfo{author}{\bibfnamefont{R.}~\bibnamefont{Reedy}},
  \bibinfo{author}{\bibfnamefont{J.}~\bibnamefont{Arnold}}, \bibnamefont{and}
  \bibinfo{author}{\bibfnamefont{J.}~\bibnamefont{Trombka}},
  \bibinfo{journal}{J. Geophys. Res.} \textbf{\bibinfo{volume}{78}},
  \bibinfo{pages}{5847} (\bibinfo{year}{1973}).

\bibitem[{\citenamefont{Boynton et~al.}(2007)\citenamefont{Boynton, Taylor,
  Reedy, Starr, Janes, Kerry, Drake, Kim, Williams, Crombie
  et~al.}}]{Boynton:2007}
\bibinfo{author}{\bibfnamefont{W.}~\bibnamefont{Boynton}},
  \bibinfo{author}{\bibfnamefont{G.}~\bibnamefont{Taylor}},
  \bibinfo{author}{\bibfnamefont{R.}~\bibnamefont{Reedy}},
  \bibinfo{author}{\bibfnamefont{R.}~\bibnamefont{Starr}},
  \bibinfo{author}{\bibfnamefont{D.}~\bibnamefont{Janes}},
  \bibinfo{author}{\bibfnamefont{K.}~\bibnamefont{Kerry}},
  \bibinfo{author}{\bibfnamefont{D.}~\bibnamefont{Drake}},
  \bibinfo{author}{\bibfnamefont{K.}~\bibnamefont{Kim}},
  \bibinfo{author}{\bibfnamefont{R.}~\bibnamefont{Williams}},
  \bibinfo{author}{\bibfnamefont{M.}~\bibnamefont{Crombie}},
  \bibnamefont{et~al.}, \bibinfo{journal}{J. Geophys. Res.}
  \textbf{\bibinfo{volume}{112}}, \bibinfo{pages}{E12S99}
  (\bibinfo{year}{2007}).

\bibitem[{\citenamefont{Skidmore and Ambrosi}(2009)}]{Skidmore:2009}
\bibinfo{author}{\bibfnamefont{M.}~\bibnamefont{Skidmore}} \bibnamefont{and}
  \bibinfo{author}{\bibfnamefont{R.}~\bibnamefont{Ambrosi}},
  \bibinfo{journal}{Adv. Space Res.} \textbf{\bibinfo{volume}{44}},
  \bibinfo{pages}{1019} (\bibinfo{year}{2009}).

\bibitem[{\citenamefont{Kiehn and Goodman}(1954)}]{Kiehn:1954}
\bibinfo{author}{\bibfnamefont{R.}~\bibnamefont{Kiehn}} \bibnamefont{and}
  \bibinfo{author}{\bibfnamefont{C.}~\bibnamefont{Goodman}},
  \bibinfo{journal}{Phys. Rev.} \textbf{\bibinfo{volume}{95}},
  \bibinfo{pages}{989} (\bibinfo{year}{1954}).

\bibitem[{\citenamefont{Day}(1956)}]{Day:1956}
\bibinfo{author}{\bibfnamefont{R.~B.} \bibnamefont{Day}},
  \bibinfo{journal}{Phys. Rev.} \textbf{\bibinfo{volume}{102}},
  \bibinfo{pages}{767} (\bibinfo{year}{1956}).

\bibitem[{\citenamefont{Cranberg and Levin}(1956)}]{Cranberg:1956}
\bibinfo{author}{\bibfnamefont{L.}~\bibnamefont{Cranberg}} \bibnamefont{and}
  \bibinfo{author}{\bibfnamefont{J.}~\bibnamefont{Levin}},
  \bibinfo{journal}{Phys. Rev.} \textbf{\bibinfo{volume}{103}},
  \bibinfo{pages}{343} (\bibinfo{year}{1956}).

\bibitem[{\citenamefont{Bredin}(1964)}]{Bredin:1964}
\bibinfo{author}{\bibfnamefont{D.}~\bibnamefont{Bredin}},
  \bibinfo{journal}{Phys. Rev.} \textbf{\bibinfo{volume}{135}},
  \bibinfo{pages}{B412} (\bibinfo{year}{1964}).

\bibitem[{\citenamefont{Rogers}(1972)}]{Rogers:1972}
\bibinfo{author}{\bibfnamefont{V.~C.} \bibnamefont{Rogers}},
  \bibinfo{journal}{Phys. Rev. C} \textbf{\bibinfo{volume}{6}},
  \bibinfo{pages}{801} (\bibinfo{year}{1972}).

\bibitem[{\citenamefont{Mellema et~al.}(1986)\citenamefont{Mellema, Finlay, and
  Dietrich}}]{Mellema:1986}
\bibinfo{author}{\bibfnamefont{S.}~\bibnamefont{Mellema}},
  \bibinfo{author}{\bibfnamefont{R.~W.} \bibnamefont{Finlay}},
  \bibnamefont{and} \bibinfo{author}{\bibfnamefont{F.~S.}
  \bibnamefont{Dietrich}}, \bibinfo{journal}{Phys. Rev. C}
  \textbf{\bibinfo{volume}{33}}, \bibinfo{pages}{481} (\bibinfo{year}{1986}).

\bibitem[{\citenamefont{Pedroni et~al.}(1988)\citenamefont{Pedroni, Howell,
  Honor\'e, Pfutzner, Byrd, Walter, and Delaroche}}]{Pedroni:1988}
\bibinfo{author}{\bibfnamefont{R.~S.} \bibnamefont{Pedroni}},
  \bibinfo{author}{\bibfnamefont{C.~R.} \bibnamefont{Howell}},
  \bibinfo{author}{\bibfnamefont{G.~M.} \bibnamefont{Honor\'e}},
  \bibinfo{author}{\bibfnamefont{H.~G.} \bibnamefont{Pfutzner}},
  \bibinfo{author}{\bibfnamefont{R.~C.} \bibnamefont{Byrd}},
  \bibinfo{author}{\bibfnamefont{R.~L.} \bibnamefont{Walter}},
  \bibnamefont{and} \bibinfo{author}{\bibfnamefont{J.~P.}
  \bibnamefont{Delaroche}}, \bibinfo{journal}{Phys. Rev. C}
  \textbf{\bibinfo{volume}{38}}, \bibinfo{pages}{2052} (\bibinfo{year}{1988}).

\bibitem[{\citenamefont{Savin et~al.}(1976)\citenamefont{Savin, Khokhlov,
  Paramonova, Chirkin, Ludin, Saraeva, and Zherebtsov}}]{Savin:1976}
\bibinfo{author}{\bibfnamefont{M.}~\bibnamefont{Savin}},
  \bibinfo{author}{\bibfnamefont{Y.}~\bibnamefont{Khokhlov}},
  \bibinfo{author}{\bibfnamefont{I.}~\bibnamefont{Paramonova}},
  \bibinfo{author}{\bibfnamefont{V.}~\bibnamefont{Chirkin}},
  \bibinfo{author}{\bibfnamefont{V.}~\bibnamefont{Ludin}},
  \bibinfo{author}{\bibfnamefont{M.}~\bibnamefont{Saraeva}}, \bibnamefont{and}
  \bibinfo{author}{\bibfnamefont{V.}~\bibnamefont{Zherebtsov}},
  \bibinfo{journal}{Sov. J. Nucl. Phys.} \textbf{\bibinfo{volume}{23}},
  \bibinfo{pages}{269} (\bibinfo{year}{1976}).

\bibitem[{\citenamefont{Sisterson and Chadwick}(2006)}]{Sisterson:2006}
\bibinfo{author}{\bibfnamefont{J.~M.} \bibnamefont{Sisterson}}
  \bibnamefont{and} \bibinfo{author}{\bibfnamefont{M.~B.}
  \bibnamefont{Chadwick}}, \bibinfo{journal}{Nucl. Instrum. Methods B}
  \textbf{\bibinfo{volume}{245}}, \bibinfo{pages}{371} (\bibinfo{year}{2006}).

\bibitem[{\citenamefont{Castaneda et~al.}(2007)\citenamefont{Castaneda,
  Gearhart, Gearhart, Sanii, Englert, Dempsey, Young, Drake, and
  Reedy}}]{Castaneda:2007}
\bibinfo{author}{\bibfnamefont{C.}~\bibnamefont{Castaneda}},
  \bibinfo{author}{\bibfnamefont{D.}~\bibnamefont{Gearhart}},
  \bibinfo{author}{\bibfnamefont{R.}~\bibnamefont{Gearhart}},
  \bibinfo{author}{\bibfnamefont{B.}~\bibnamefont{Sanii}},
  \bibinfo{author}{\bibfnamefont{P.}~\bibnamefont{Englert}},
  \bibinfo{author}{\bibfnamefont{J.}~\bibnamefont{Dempsey}},
  \bibinfo{author}{\bibfnamefont{J.}~\bibnamefont{Young}},
  \bibinfo{author}{\bibfnamefont{D.}~\bibnamefont{Drake}}, \bibnamefont{and}
  \bibinfo{author}{\bibfnamefont{R.}~\bibnamefont{Reedy}},
  \bibinfo{journal}{Nucl. Instrum. Methods B} \textbf{\bibinfo{volume}{260}},
  \bibinfo{pages}{508} (\bibinfo{year}{2007}).

\bibitem[{\citenamefont{Lachkar et~al.}(1974)\citenamefont{Lachkar, Sigaud,
  Patin, and Haouat}}]{Lachkar:1974}
\bibinfo{author}{\bibfnamefont{J.}~\bibnamefont{Lachkar}},
  \bibinfo{author}{\bibfnamefont{J.}~\bibnamefont{Sigaud}},
  \bibinfo{author}{\bibfnamefont{Y.}~\bibnamefont{Patin}}, \bibnamefont{and}
  \bibinfo{author}{\bibfnamefont{G.}~\bibnamefont{Haouat}},
  \bibinfo{journal}{Nucl. Sci. Eng.} \textbf{\bibinfo{volume}{55}},
  \bibinfo{pages}{168} (\bibinfo{year}{1974}).

\bibitem[{\citenamefont{Xiamin et~al.}(1982)\citenamefont{Xiamin, Ronglin,
  Jinqiang, Yongshun, and Dazhao}}]{Xiamin:1982}
\bibinfo{author}{\bibfnamefont{S.}~\bibnamefont{Xiamin}},
  \bibinfo{author}{\bibfnamefont{S.}~\bibnamefont{Ronglin}},
  \bibinfo{author}{\bibfnamefont{X.}~\bibnamefont{Jinqiang}},
  \bibinfo{author}{\bibfnamefont{W.}~\bibnamefont{Yongshun}}, \bibnamefont{and}
  \bibinfo{author}{\bibfnamefont{D.}~\bibnamefont{Dazhao}},
  \bibinfo{journal}{Chin. J. Nucl. Phys.} \textbf{\bibinfo{volume}{4}},
  \bibinfo{pages}{120} (\bibinfo{year}{1982}).

\bibitem[{\citenamefont{Heusser}(1996)}]{Heusser:1996}
\bibinfo{author}{\bibfnamefont{G.}~\bibnamefont{Heusser}},
  \bibinfo{journal}{Nucl. Instrum. Methods A} \textbf{\bibinfo{volume}{369}},
  \bibinfo{pages}{539 } (\bibinfo{year}{1996}).

\bibitem[{\citenamefont{Kudryavtsev et~al.}(2003)\citenamefont{Kudryavtsev,
  Spooner, and McMillan}}]{Kudryavtsev:2003}
\bibinfo{author}{\bibfnamefont{V.}~\bibnamefont{Kudryavtsev}},
  \bibinfo{author}{\bibfnamefont{N.}~\bibnamefont{Spooner}}, \bibnamefont{and}
  \bibinfo{author}{\bibfnamefont{J.}~\bibnamefont{McMillan}},
  \bibinfo{journal}{Nucl. Instrum. Methods A} \textbf{\bibinfo{volume}{505}},
  \bibinfo{pages}{688} (\bibinfo{year}{2003}).

\bibitem[{\citenamefont{Galbiati and Beacom}(2005)}]{Galbiati:2005}
\bibinfo{author}{\bibfnamefont{C.}~\bibnamefont{Galbiati}} \bibnamefont{and}
  \bibinfo{author}{\bibfnamefont{J.~F.} \bibnamefont{Beacom}},
  \bibinfo{journal}{Phys. Rev. C} \textbf{\bibinfo{volume}{72}},
  \bibinfo{pages}{025807} (\bibinfo{year}{2005}).

\bibitem[{\citenamefont{Schr\"oder et~al.}(1974)\citenamefont{Schr\"oder,
  Jahnke, Lindenberger, R\"oschert, Engfer, and Walter}}]{Schroeder:1974}
\bibinfo{author}{\bibfnamefont{W.}~\bibnamefont{Schr\"oder}},
  \bibinfo{author}{\bibfnamefont{U.}~\bibnamefont{Jahnke}},
  \bibinfo{author}{\bibfnamefont{K.}~\bibnamefont{Lindenberger}},
  \bibinfo{author}{\bibfnamefont{G.}~\bibnamefont{R\"oschert}},
  \bibinfo{author}{\bibfnamefont{R.}~\bibnamefont{Engfer}}, \bibnamefont{and}
  \bibinfo{author}{\bibfnamefont{H.}~\bibnamefont{Walter}},
  \bibinfo{journal}{Z. Phys. A} \textbf{\bibinfo{volume}{268}},
  \bibinfo{pages}{57} (\bibinfo{year}{1974}).

\bibitem[{\citenamefont{Agafonova et~al.}(2009)\citenamefont{Agafonova,
  Boyarkin, Dadykin, Dobrynina, Enikeev, Kuznetsov, Malgin, Ryazhskaya, Ryasny,
  Yakushev et~al.}}]{Agafonova:2009}
\bibinfo{author}{\bibfnamefont{N.}~\bibnamefont{Agafonova}},
  \bibinfo{author}{\bibfnamefont{V.}~\bibnamefont{Boyarkin}},
  \bibinfo{author}{\bibfnamefont{V.}~\bibnamefont{Dadykin}},
  \bibinfo{author}{\bibfnamefont{E.}~\bibnamefont{Dobrynina}},
  \bibinfo{author}{\bibfnamefont{R.}~\bibnamefont{Enikeev}},
  \bibinfo{author}{\bibfnamefont{V.}~\bibnamefont{Kuznetsov}},
  \bibinfo{author}{\bibfnamefont{A.}~\bibnamefont{Malgin}},
  \bibinfo{author}{\bibfnamefont{O.}~\bibnamefont{Ryazhskaya}},
  \bibinfo{author}{\bibfnamefont{V.}~\bibnamefont{Ryasny}},
  \bibinfo{author}{\bibfnamefont{V.}~\bibnamefont{Yakushev}},
  \bibnamefont{et~al.}, \bibinfo{journal}{Bull. Russ. Acad. Sci. Phys.}
  \textbf{\bibinfo{volume}{73}}, \bibinfo{pages}{628} (\bibinfo{year}{2009}).

\bibitem[{\citenamefont{Allison et~al.}(2006)\citenamefont{Allison, Amako,
  Apostolakis, Araujo, Dubois, Asai, Barrand, Capra, Chauvie, Chytracek
  et~al.}}]{GEANT4}
\bibinfo{author}{\bibfnamefont{J.}~\bibnamefont{Allison}},
  \bibinfo{author}{\bibfnamefont{K.}~\bibnamefont{Amako}},
  \bibinfo{author}{\bibfnamefont{J.}~\bibnamefont{Apostolakis}},
  \bibinfo{author}{\bibfnamefont{H.}~\bibnamefont{Araujo}},
  \bibinfo{author}{\bibfnamefont{P.}~\bibnamefont{Dubois}},
  \bibinfo{author}{\bibfnamefont{M.}~\bibnamefont{Asai}},
  \bibinfo{author}{\bibfnamefont{G.}~\bibnamefont{Barrand}},
  \bibinfo{author}{\bibfnamefont{R.}~\bibnamefont{Capra}},
  \bibinfo{author}{\bibfnamefont{S.}~\bibnamefont{Chauvie}},
  \bibinfo{author}{\bibfnamefont{R.}~\bibnamefont{Chytracek}},
  \bibnamefont{et~al.}, \bibinfo{journal}{IEEE Transactions on Nuclear Science}
  \textbf{\bibinfo{volume}{53}}, \bibinfo{pages}{270} (\bibinfo{year}{2006}).

\bibitem[{\citenamefont{Heisinger et~al.}(2002)\citenamefont{Heisinger, Lal,
  Jull, Kubik, Ivy-Ochs, Knie, and Nolte}}]{Heisinger:2002}
\bibinfo{author}{\bibfnamefont{B.}~\bibnamefont{Heisinger}},
  \bibinfo{author}{\bibfnamefont{D.}~\bibnamefont{Lal}},
  \bibinfo{author}{\bibfnamefont{A.}~\bibnamefont{Jull}},
  \bibinfo{author}{\bibfnamefont{P.}~\bibnamefont{Kubik}},
  \bibinfo{author}{\bibfnamefont{S.}~\bibnamefont{Ivy-Ochs}},
  \bibinfo{author}{\bibfnamefont{K.}~\bibnamefont{Knie}}, \bibnamefont{and}
  \bibinfo{author}{\bibfnamefont{E.}~\bibnamefont{Nolte}},
  \bibinfo{journal}{Earth Planet. Sci. Lett.} \textbf{\bibinfo{volume}{200}},
  \bibinfo{pages}{357} (\bibinfo{year}{2002}).

\bibitem[{\citenamefont{Mannhart}(1987)}]{Mannhart:1987}
\bibinfo{author}{\bibfnamefont{W.}~\bibnamefont{Mannhart}},
  \bibinfo{type}{Technical Report Series} \bibinfo{number}{273},
  \bibinfo{institution}{IAEA}, \bibinfo{address}{Vienna}
  (\bibinfo{year}{1987}).

\end{thebibliography}

\end{document}